\documentclass[ACS,STIX1COL]{WileyNJD-v2}
\articletype{ORIGINAL ARTICLE}

\received{xxx}
\revised{xxx}
\accepted{xxx}

\usepackage{bm}

\begin{document}

\title{Neutralization dynamics of slow highly charged ions passing through graphene nanoflakes--an embedding self-energy approach}

\author[1]{K. Balzer}
\author[2]{M. Bonitz}

\authormark{K. Balzer \textsc{et al}}

\address[1]{\orgdiv{Rechenzentrum},
\orgname{Christian-Albrechts-Universit\"at zu Kiel}, \orgaddress{\state{24098 Kiel}, \country{Germany}}}

\address[2]{\orgdiv{Institut f\"ur Theoretische Physik und Astrophysik}, \orgname{Christian-Albrechts-Universit\"at zu Kiel}, \orgaddress{\state{Leibnizstra{\ss}e 15, 24098 Kiel}, \country{Germany}}}

%% \address[3]{\orgdiv{Org Division}, \orgname{Org Name}, \orgaddress{\state{State name}, \country{Country name}}}

\corres{*\email{balzer@rz.uni-kiel.de}}

%% \presentaddress{This is sample for present address text this is sample for present address text}

\abstract{We study the time-dependent neutralization of a slow highly charged ion that  penetrates a hexagonal hollow-centred graphene nanoflake. To compute the ultrafast charge transfer dynamics, we apply an effective Hubbard nanocluster model and use the method of nonequilibrium Green functions (NEGF) in conjunction with an embedding self-energy scheme which allows one to follow the temporal changes of the number of electrons in the nanoflake. We perform extensive  simulations of the charge transfer dynamics for a broad range of ion charge states and impact velocities. The results are used to put forward a simple semi-analytical model of the neutralization dynamics that is in very good agreement with transmission experiments, in which highly charged xenon ions pass through sheets of single-layer graphene.}


\maketitle

%% \footnotetext{\textbf{Abbreviations:} ANA, anti-nuclear antibodies; APC, antigen-presenting cells; IRF, interferon regulatory factor}

\section{Introduction}\label{sec:1}

The interaction of heavy charged particles with matter is of fundamental importance in various fields of physics. Particularly, the collision of ions with surfaces of solids is of high current interest both, theoretically and for technological applications in plasmas and plasma-surface interactions~\cite{Bonitz_FCSE_2019}. The ion-induced excitation and ionization of electrons in the surface material as well as the capture of electrons by the ion (which changes the charge state of the ion) is crucial for the electrostatic potential at the surface and in the plasma sheath region. The theoretical description of charge transfer processes in the plasma near the surface region has been the subject of intense investigations, for recent studies and further references, see, e.g., Refs.~\cite{Marbach2012,Pamperin_Bronold_PRB_2015}.  
In recent years, a number of studies was devoted to the energy loss (stopping power) of charged particles in strongly correlated solids,  nanostructures and single-layered materials, such as graphene \cite{Zhao_2014, balzer-schluenzen_PRB_2016,Balzer_PRL_2018}. 

Another topic that caught attention has been  the irradiation of solids by highly charged ions (HCIs)~\cite{Aumayr2003,Facsko_2009}, in particular, the impact of HCIs on single-layer graphene \cite{Gruber2016}.
%
%%
%Due to the loss of many or even most of their bound electrons, 
Due to their high charge these ions carry a large amount of potential energy~\cite{Aumayr2003,Facsko_2009} and, depending on
their kinetic energy, reveal manifold and complex interactions with surfaces of solids. For slow HCIs with typical velocities of $v\sim\!10^5\,$m/s, i.e., with velocities less than the Bohr velocity [the electron velocity in the first Bohr orbit], surface transmission times are of the order of a few femtoseconds and are hence comparable to the time scale of the electronic motion in the solid. In this regime, the exposure of surfaces to individual or ensembles of highly charged ions allows for interesting nanoscale applications including surface (re)structuring and modification~\cite{Kentsch_2004,Aumayr_2008,Facsko_2009,Ritter_2013} and surface diagnostics~\cite{Schenkel_1999}.
%%
%% s=v*t --> t=s/v --> v=1.0*10^5 m/s, s=20.0*10^(-10) m --> t=s/v=20.0*10^(-10)/(1.0*10^5) s = 20.0*10^(-15) s = 20.0 fs 
%%
Moreover, upon ion impact, the extraction of electrons from the surface can lead to partial or complete neutralization of the ion~\cite{Gruber2016} as well as to secondary electron emission~\cite{Meissl_2008}. The loss of charge carriers in the surface, together with the deposition of potential energy, can ultimately culminate in the  destruction of surface domains in a Coulomb explosion-like sputtering scenario~\cite{Tona_2005}. How exactly a surface and its electronic structure reacts to the impact of a highly charged ion depends on multiple factors, in particular on the ion's charge state, the ion velocity and the material properties. Interesting questions, which have recently been addressed in that regard, concern, for example, the formation of correlated two-particle states (doublons) during ion impact~\cite{Balzer_PRL_2018,schluenzen-balzer_CPP_2019,bonitz-pssb2019} or the influence of the band gap on the charge exchange~\cite{Creutzburg2020}. Of interest is also the impact of multiple ions and possible correlations of their individual effects. This requires to go beyond linear response theory and to perform a non-adiabatic time-dependent correlated simulation, see Refs. \cite{Balzer_PRL_2018,bonitz-pssb2019}.
In these works a combined Nonequilibrium Green Functions(NEGF)-Ehrenfest approach\cite{stefanucci13_cambridge,balzer-book} was developed ~\cite{schluenzen-balzer_CPP_2019,balzer-schluenzen_PRB_2016} that focused on ion stopping and energy loss spectra (electronic stopping power) for closed surface models. However, the processes related to charge transfer were not included in these simulations.

The goal of the present article, is to extend these time-resolved NEGF-Ehrenfest simulations to include charge transfer processes. As a test case, we simulate the electron emission from two-dimensional hexagonal hollow-centered graphene nanoflakes (GNFs) during the impact of slow HCIs. In the course of this, neutralizing charge transfer to the ion is described via an embedding self-energy scheme, which mimics the formation of a highly excited hollow atom in front of the GNF surface. The approach is capable to resolve the ultrafast electronic response of the nanoflake at accurate exit charge states of the ion, while the de-excitation and recombination processes inside the ion are not explicitly considered. 
The article is organized as follows. Section~\ref{sec:2} outlines the effective Hubbard-type model that is used to describe the graphene nanoflake, including electronic correlations, and its interaction with the ion. Further, we here motivate the spatio-temporal form of the charge transfer amplitude. Section~\ref{sec:3} presents the theory, derives the central Keldysh-Kadanoff-Baym equations of the GNF and introduces the embedding self-energy, which couples ion and surface states. Results of the NEGF calculations are compiled in Sec.~\ref{sec:4}, and the conclusions are summarized in Sec.~\ref{sec:5}. For the system of units used throughout the article, see below Eq.~(\ref{eq:ham}).

\section{Model}\label{sec:2}

\begin{figure}
\begin{center}
\includegraphics[width=0.975\textwidth]{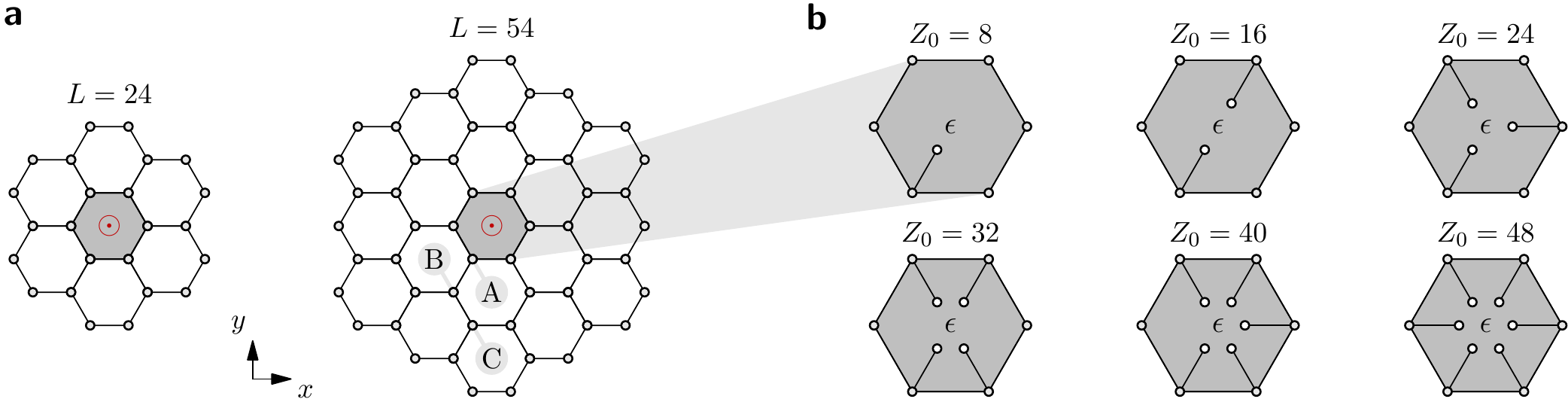}
\end{center}
\caption{(a)~Hexagonal hollow-centered graphene nanoflakes (GNF) with a finite number of honeycombs and $L=24$ and $54$ lattice sites. The red dot ($\odot$) in the center of the GNFs indicates the impact point of the  approaching ion. (b)~Charge transfer couplings considered in the present model, Eq. (\ref{eq:ham}). Depending on the initial ion charge state $Z_0$, only a selected set of central GNF sites $k\in M$  participate in the resonant transfer of electrons at an effective single-particle energy $\epsilon_k=\epsilon$.}
\label{fig:gnf}
\end{figure}

The electrons in the covalent $\sigma$- and $\pi$-bonds of the sp$^2$-hybridized carbon atoms of the GNF are described in terms of an effective Hubbard nanocluster  model with four equal decoupled bands at half-filling, cf.~Fig.~\ref{fig:gnf}\!\!\!a for the lattice structure of GNFs with $L=24$ and $54$ sites. Using a nearest-neighbor hopping $J$, a local Coulomb repulsion $U$ and an on-site energy $E$, the overall Hamiltonian is given by $\hat{H}(t)=4\hat{H}_\textup{s}(t)$, with the single-band contribution
\begin{align}
\label{eq:ham}
\hat{H}_\textup{s}(t)=&\sum_{i\sigma}\underbrace{\left\{ W_{i}(Z_0;{\bf S}(t))+E\right\}}_{=\,E_i({\bf S}(t))}\,\hat{c}_{i\sigma}^{\dagger}\hat{c}_{i\sigma}+\sum_{\langle ij\rangle,\sigma}\left\{ W_{\langle ij\rangle}(Z_0;{\bf S}(t))-J\right\}\,\hat{c}_{i\sigma}^{\dagger}\hat{c}_{j\sigma}+U\sum_{i}\left(\hat{n}_{i\uparrow}-\frac{1}{2}\right)\left(\hat{n}_{i\downarrow}-\frac{1}{2}\right)\nonumber\\&+\sum_{k\in M,\sigma}\gamma_{k}[{\bf S}(t)]\left(\hat{a}_{k\sigma}^\dagger\hat{c}_{k\sigma}+\hat{c}_{k\sigma}^\dagger\hat{a}_{k\sigma}\right)+\sum_{k\in M,\sigma}\epsilon_k\,\hat{a}_{k\sigma}^\dagger\hat{a}_{k\sigma}\,.
\end{align}
The operator $\hat{c}^\dagger_{i\sigma}$ ($\hat{c}_{i\sigma}$) creates (annihilates) an electron with spin $\sigma\in\{\uparrow,\downarrow\}$ on site $i$ of the GNF oriented in the $x-y$-plane, $\hat{n}_{i\sigma}=\hat{c}^\dagger_{i\sigma}\hat{c}_{i\sigma}$ denotes the local electron density, $\langle ij\rangle$ indicates nearest-neighbor contributions, and the matrix elements $W_{i}(Z_0;{\bf S}(t))$ and $W_{\langle ij\rangle}(Z_0;{\bf S}(t))$ describe the time-dependent Coulomb interaction of the electrons with a highly charged ion, which has initial charge state $Z_0$ and follows a normal incidence trajectory ${\bf S}(t)=(0,0,z(t)=v_zt)$ with a constant velocity $v_z$, cf.~Sec.~\ref{sec:2.1}. Furthermore, the last two terms in Eq.~(\ref{eq:ham}) allow for charge transfer from a set of selected lattice sites $k\in M$ onto initially unoccupied energy levels $\epsilon_k$ of the external ion, where $a^\dagger_{k\sigma}$ ($a_{k\sigma}$) are the corresponding creation (annihilation) operators, cf.~Sec.~\ref{sec:2.2} for the discussion of the trajectory-dependent charge transfer amplitude $\gamma_k[{\bf S}(t)]$ and the energies $\epsilon_k$. Depending on the charge state $Z_0$, we couple per Hubbard band only a few of the central GNF sites to the ion, cf. Fig.~\ref{fig:gnf}\!\!\!b. For a more detailed discussion on this choice, see end of Sec.~\ref{sec:3}.

Throughout the article, we measure energies in units of
%% $J_0=\hbar^2/(m a_0^2)=3.778996269$\,eV
$J_0=\hbar^2/(m a_0^2)\approx3.78$\,eV
and times in units of
%% $t_0=\hbar/J_0=0.174176461$\,fs, 
$t_0=\hbar/J_0\approx0.17$\,fs, 
where $m$ denotes the electron mass, and $a_0=1.42$\,{\AA} is the molecular carbon-carbon bond length. The ion velocity $v_z$ is given in units of
%% $v_0=a_0/t_0=0.815265157$\,nm/fs.
$v_0=a_0/t_0\approx0.82$\,nm/fs$=8.2 \cdot 10^5$\,m/s.

\subsection{Kinetic, potential, and interaction energy\label{sec:2.1}}
To determine the single-particle matrix elements of the kinetic and potential energies in Eq.~(\ref{eq:ham}), we use localized atomic orbitals of Gaussian form, $\varphi_{i\sigma}({\bf r})=\frac{1}{\pi^{3/4}\mu^{3/2}}\textup{e}^{-({\bf r}-{\bf R}_i)^2/(2\mu^2)}$, with width $\mu=\mu_0 a_0$,
and assume the potential close to a lattice site with coordinate ${\bf R}_i$ to be
$v_i({\bf r})=-V\textup{e}^{-({\bf r}-{\bf R}_i)^2/(2\nu^2)}$, with $\nu=\nu_0 a_0$ and $V=V_0 J_0$. Below we will use the parameters $\mu_0$, $\nu_0$ and $V_0$ as free parameters of the model, their values for the case of graphene are specified in Eqs.~(\ref{eq:condition1}) and (\ref{eq:condition2}).

For sites located at positions ${\bf R}_i=(0,0,0)$ and ${\bf R}_j=(a_0,0,0)$, the kinetic part, $T_{\langle ij\rangle}$, of the nearest-neighbor hopping matrix element, $-J=T_{\langle ij\rangle}+V_{\langle ij\rangle}$, is given by
\begin{align}
\label{eq:tij}
T_{\langle ij\rangle}&=-\frac{\hbar^2}{2m}\int \textup{d}^3r\,\phi^*_{i\sigma}({\bf r})\nabla^2\phi_{j\sigma}({\bf r})
=-\frac{1}{8\mu_0^4}(1-6\mu_0^2)\,\textup{e}^{-1/(4\mu^2)}J_0\,,
\end{align}
whereas, for the potential part, $V_{\langle ij\rangle}$, we obtain
\begin{align}
\label{eq:vij}
 V_{\langle ij\rangle}&=\int \textup{d}^3r\,\phi^*_{i\sigma}({\bf r})\left\{v_i({\bf r})+v_j({\bf r})\right\}\phi_{j\sigma}({\bf r})=-\frac{4\sqrt{2}\,\nu_0^3\textup{e}^{-(\mu_0^2+\nu_0^2)/(2\mu_0^2(\mu_0^2+2\nu_0^2))}V_0}{(\mu_0^2+2\nu_0^2)^{3/2}}J_0\,.
\end{align}
For the on-site energy $E=T_{i}+V_{i}$, we have
\begin{align}
\label{eq:ti}
T_{i}&=-\frac{\hbar^2}{2m}\int \textup{d}^3r\,\phi^*_{i\sigma}({\bf r})\nabla^2\phi_{i\sigma}({\bf r})=\frac{3}{4\mu_0^2}J_0\,,\\
\label{eq:vi}
V_{i}&=\int \textup{d}^3r\,\phi^*_{i\sigma}({\bf r})v_i({\bf r})\phi_{i\sigma}({\bf r})=-\frac{2\sqrt{2} \nu_0^3V_0}{(\mu_0^2+2\nu_0^2)^{3/2}}J_0\,.
\end{align}
Furthermore, the time-dependent Coulomb potential generated by the ion,
\begin{align}
\label{eq:wpotential}
    w[Z_0;{\bf r}-{\bf S}(t)]&=-\frac{Z_0 e^2}{4\pi\epsilon_0}|{\bf r}-{\bf S}(t)|^{-1}=-Z_0W_0a_0 |{\bf r}-{\bf S}(t)|^{-1}\;,
\end{align}
with electron charge $e$, vacuum permittivity $\epsilon_0$ and 
%% $W_0=e^2/(4\pi\epsilon_0a_0)=10.14059981$\,eV ($W_0/J_0=2.683410907$),
$W_0=e^2/(4\pi\epsilon_0a_0)\approx10.14$\,eV ($W_0/J_0\approx2.68$),
leads to a modification of the on-site energy and the nearest-neighbor hopping. The corresponding matrix elements are treated as follows:
\begin{align}
\label{eq:wi}
W_{i}[Z_0;{\mathbf{S}}(t)]&=\int \textup{d}^3r\,\phi^*_{i\sigma}({\bf r})w[Z_0;{\bf r}-{\bf S}(t)]\phi_{i\sigma}({\bf r})\approx w[Z_0;{\bf R}_i-{\bf S}(t)]\,,\\
\label{eq:wij}
W_{\langle ij\rangle}[Z_0;{\mathbf{S}}(t)]&=\int \textup{d}^3r\,\phi^*_{i\sigma}({\bf r})w[Z_0;{\bf r}-{\bf S}(t)]\phi_{j\sigma}({\bf r})\approx\frac{\lambda}{2}\left(w[Z_0;{\bf R}_i-{\bf S}(t)]+w[Z_0;{\bf R}_j-{\bf S}(t)]\right)\,,
\end{align}
where $\lambda$ denotes the overlap integral
$\lambda=\int_{\langle ij\rangle} \textup{d}^3r\,\phi^*_{i\sigma}({\bf r})\phi_{j\sigma}({\bf r})=\textup{e}^{-1/(4\mu_0^2)}$, defined with respect to nearest neighbor sites.

To determine the free parameters $\mu_0$, $\nu_0$ and $V_0$ in Eqs.~(\ref{eq:tij})-(\ref{eq:vi}), we require 
\begin{align}
\label{eq:condition1}
-J=T_{\langle ij\rangle}+V_{\langle ij\rangle}&=-2.8\,\textup{eV},\\
\label{eq:condition2}
E=T_{i}+V_{i}=-W_\textup{f}&=-4.6\,\textup{eV},
\end{align}
where
%% $J=2.8$\,eV ($J/J_0=0.740937487$)
$J=2.8$\,eV ($J/J_0\approx0.74$)
is the typical carbon-carbon hopping energy in graphene~\cite{Schueler2013}, and $W_\textup{f}=4.6$\,eV
%% ($W_\textup{f}/J_0=1.217254443$) 
($W_\textup{f}/J_0\approx1.22$) 
is the work function of
single-layer graphene~\cite{Yu2009}. This leads to a nonlinear system of equations, which has a reasonable solution for
%% $\mu_0=0.3652468$,  $\nu_0=\lambda=0.15351$ and $V_0=51.274$. 
$\mu_0=0.365$,  $\nu_0=\lambda=0.153$ and $V_0=51.274$. 
Finally, following Ref.~\cite{Schueler2013}, we set the on-site Coulomb repulsion $U$ in Hamiltonian (\ref{eq:ham}) to $U=1.6J=4.48$\,eV
%% ($U/J_0=1.18549998$),
($U/J_0\approx1.19$)
and, neglecting electron-electron correlations, use the Hartree approximation,
\begin{align}
\label{eq:hartree}
    U\left(\hat{n}_{i\uparrow}-\frac{1}{2}\right)\left(\hat{n}_{i\downarrow}-\frac{1}{2}\right)\approx U\sum_{\sigma}\left(\langle\hat{n}_{i\bar{\sigma}}\rangle_\textup{s}(t)-\frac{1}{2}\right)\hat{c}^\dagger_{i\sigma}\hat{c}_{i\sigma}\,,
\end{align}
where $\langle\hat{n}_{i\uparrow}\rangle_\textup{s}(t)=\langle\hat{n}_{i\downarrow}\rangle_\textup{s}(t)$ denotes the expectation value of the local electron density with respect to $\hat{H}_\textup{s}(t)$. The computational approach outlined in Sec.~\ref{sec:3} is, however, not limited to a mean-field treatment and can be straightforwardly extended to include electron-electron correlations via the many-body self-energy, cf.~Eq.~(\ref{eq:kbefinal}) in Sec.~\ref{sec:3}. 

\subsection{Charge transfer amplitude\label{sec:2.2}}

To get an idea of how the charge transfer amplitude $\gamma_k[\bf{S}(t)]$ of Eq.~(\ref{eq:ham}) varies as function of time and ion-surface distance $z(t)$, respectively, we start from the energy spectrum of a 
%hydrogenic, i.e., 
hydrogen-like ion with charge (atomic number) $Z$, which is given by
\begin{align}
\label{eq:Ehydrogenic}
    E_n(Z)=-E_\textup{R}\frac{Z^2}{n^2}\,,
\end{align}
with Rydberg energy $E_\textup{R}=1$\,Ry
%% $=13.605693122$\,eV ($E_\textup{R}=3.600345741 J_0$) and
$\approx13.61$\,eV ($E_\textup{R}\approx3.60 J_0$) and
$n=1,2,3,\ldots,\infty$.
Equation~(\ref{eq:Ehydrogenic}) represents a very simple approximation for the energy level structure of the highly excited states of the impacting ion as function of the charge state $Z$. The quantity $\textup{d}n/\textup{d}E$ is used in the following to characterize the number of final states that can be reached by an emitted electron during the neutralizing charge transfer. The number of final states $dn$ per energy interval $dE$ is obtained by inversion of Eq.~(\ref{eq:Ehydrogenic}),
%%leads to 
\begin{align}
\label{eq:doshydrogenic}
    \frac{\textup{d}n}{\textup{d}E}(Z,E)=
    %\frac{\textup{d}}{\textup{d}E}\left(Z\sqrt{-\frac{E_\textup{R}}{E}}\right)=
    \frac{Z}{2}\left(-\frac{E_\textup{R}}{E}\right)^{-1/2}\! \left(\frac{E_\textup{R}}{E^2}\right)\,.
\end{align}
For an electron with energy $E_k$ on site $k\in M$ of the graphene nanoflake, the probability for resonant charge transfer onto the ion will then be proportional to Eq.~(\ref{eq:doshydrogenic}) and proportional to some distance-dependent matrix element $\Omega_k$. Parametrizing all quantities via the ion trajectory, $\textbf{S}(t)$, we define 
\begin{align}
\label{eq:gammapropto}
    \gamma_k[\mathbf{S}(t)]=\gamma \,\cdot\,Z[\mathbf{S}(t)] \,\cdot\,\underbrace{\frac{1}{2}\left(-\frac{E_\textup{R}}{E_k}\right)^{-1/2}\! \left(\frac{E_\textup{R}}{E_k^2}\right)E_\textup{R}}_{\displaystyle=\frac{\textup{d}n}{\textup{d}E}(Z,E_k)\,\frac{E_\textup{R}}{Z}}\,\cdot\,\Omega_k[\mathbf{R}_k-\mathbf{S}(t)]\,,
%%    \gamma_k[\mathbf{S}(t)]=\gamma \,\cdot\,Z[\mathbf{S}(t)] \,\cdot\,\frac{\textup{d}n}{\textup{d}E}\left[Z[\mathbf{S}(t)],E_k[\mathbf{S}(t)]\right]\frac{E_\textup{R}}{Z[\mathbf{S}(t)]}\,\cdot\,\Omega_k[\mathbf{R}_k-\mathbf{S}(t)]\,,
\end{align}
where $E_k[\mathbf{S}(t)]=w[Z_0;\mathbf{R}_k-\mathbf{S}(t)]-W_\textup{f}\,$, (cf.~Sec~\ref{sec:2.1}), and $\gamma=\gamma_0J_0$ is an overall proportionality factor. Equation~(\ref{eq:gammapropto}) describes the time- and distance-dependent probability for an electron to undergo a resonant transition from a site $k$ of the graphene nanoflake to the ion (or vice versa) and thereby contains information on, both, the nature of the highly charged ion, as well as of the electronic structure of the initial and final states via the overlap matrix element $\Omega_k$. 

From experiments it is known that the charge transfer amplitude (\ref{eq:gammapropto}) is peaked in front of the surface, at a characteristic ion-surface distance. For this reason, we choose an analytical form for $\gamma_k[\mathbf{S}(t)]$ which, on the one hand, directly includes such a maximum and,
on the other hand, is 
 still sufficiently general. A reasonable choice is a Gaussian which, aside from the prefactor $\gamma_0$, involves only two further open parameters, $z_\textup{res}$ and $d_\textup{w}$,
\begin{align}
\label{eq:gammafinal}
    \gamma_k[{\mathbf S}(t)]=\gamma_0\exp\!\left(-\frac{(z(t)+z_\textup{res})^2}{2d_\textup{w}^2}\right)J_0\,,
\end{align}
where $z_\textup{res}$ marks the point of resonance, i.e., the ion-surface distance (point in time) where the charge transfer probability has its maximum,
%% determines the temporal (spatial) location of the maximum, 
and $d_\textup{w}$ sets a characteristic spatial (temporal) width over which the charge transfer probability is nonzero.
In the following, we discuss how the parameters $z_\textup{res}$ and $d_\textup{w}$ can be determined by comparing Eqs.~(\ref{eq:gammapropto}) and (\ref{eq:gammafinal}). To this end, we  have to make reasonable assumptions for the dependencies of the charge state $Z$ as well as the matrix element $\Omega_k$ in Eq.~(\ref{eq:gammapropto}) on the ionic position (i.e. on time). As a further simplification, we will take into account charge transfer only from lattice sites which belong to the GNF's innermost honeycomb and have thus the lateral distance $a_0$ to the impact point of the ion,  cf.~Fig.~\ref{fig:gnf}\!\!\!.
%
%\textcolor{red}{bis hierhin ist alles gut. Jetzt muessten Z und Omega eingeführt werden, Fig. 2 diskutiert werden.}
%
%\textcolor{red}{Bis jetzt ist ja alles noch allgemein. Jetzt werden die Näherungen eingeführt, die wir am Anfang selbstbewusst motivieren müssen. Dass die Kurve fuer gamma gepeakt ist, weiss man aus dem Experiment. \textbf{Vielleicht sollte man damit (also Glg. 18) beginnen.}
%Daher wählen wir eine analytische Form (Gauss bietet sich immer an), die immer noch recht allgemein ist, aber nur noch eine kleine Zahl (drei) freier Parameter erfordert. Der nächste Schritt wäre zu motivieren, wie die input-Parameter für gamma in 14 sinnvoll genähert werden können. Hier bereits auf Fig. 3 hinweisen.) Alle Näherungen schon hier einzuführen, vermeidet den Eindruck, dass wir immer neue Approximationen hinzunehmen (noch mal vor 18).}
%
For the time dependence of the ionic charge state, $Z$, we use a functional form which in a step-like but continuous behavior drops off from the initial charge state $Z_0$ and approaches zero before the ion passes though the graphene layer,
%% \textcolor{red}{auch das etwas motivieren. Dass eine Art Stufe für $Z$ rauskommen muss und gamma $\sim$ die Ableitung davon ist, könnte man verstehen. Bleibt dann die Frage, wie die beiden peak-Positionen und Breiten zusammenhängen - damit die "Willkür" weiter reduziert wird.}
%the approximation
\begin{align}
\label{eq:z}
    Z[{\bf S}(t)]=\frac{Z_0}{\textup{e}^{[z(t)+z^*]/\eta}+1}\,,
\end{align}
which leads to an approximately  constant neutralization rate around $z(t)=z^*$ and ensures a smooth behavior around the onset of neutralization as well as during the termination of the charge transfer. The parameters $\eta=\eta_0a_0$ and $z^*=z_0^*a_0$ in Eq.~(\ref{eq:z}) are chosen such that neutralization starts visibly around a few lattice spacings ($\sim\!5$-$7$\,\AA), in front of the GNF surface, and terminates at around $z=-a_0$, which is motivated by the experimental results of Ref.~\cite{Gruber2016}. Further, for the overlap matrix element, $\Omega_{k^*}$, 
where $k^*$ indicates a site index on the central honeycomb, a reasonable choice is again a Gaussian,
\begin{align}
\label{eq:omega}
    \Omega_{k^*}[{\bf R}_k-{\bf S}(t)]=\textup{e}^{-z^2(t)/(4\zeta^2)}\,,
\end{align}
with $\zeta=\zeta_0a_0$, which exhibits a maximum when the ion penetrates the $x$-$y$-plane of the graphene nanoflake. 

\begin{figure}
\begin{center}
\includegraphics[width=0.485\textwidth]{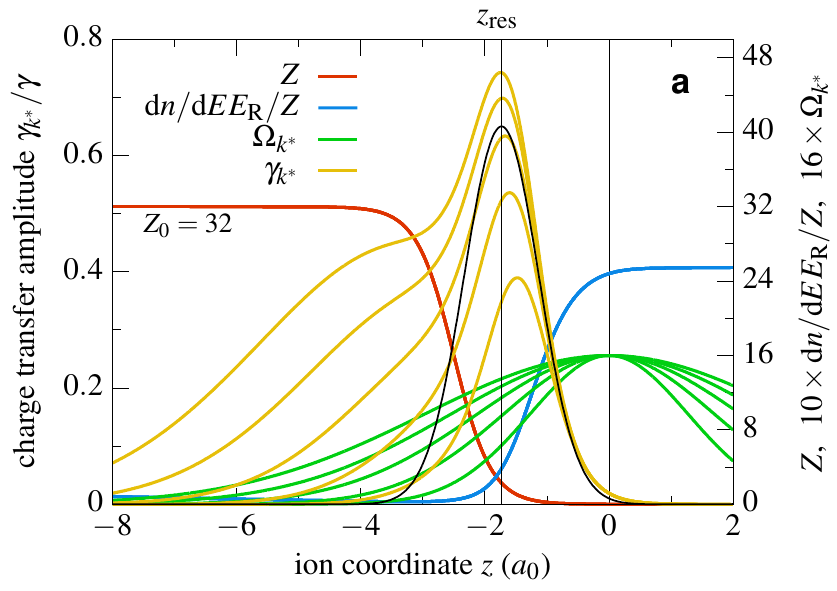}\hspace{0.015\textwidth}
\includegraphics[width=0.485\textwidth]{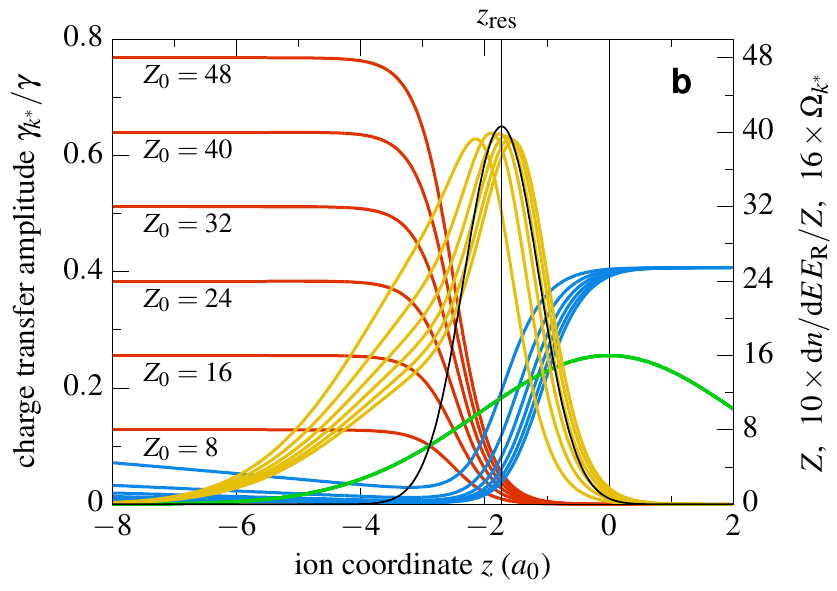}
\end{center}
\caption{Analysis of the charge transfer amplitude (orange curves). Evaluation of Eq.~(\ref{eq:gammapropto}) for ions with different initial charge states $Z_0$ (as indicated in the figure), using the assumptions (\ref{eq:z}) and  (\ref{eq:omega}) with $\eta_0=0.3$ and $z_0^*=-2.5$. In panel~(a), the parameter $\zeta_0$ is varied between $\zeta_0=0.9$ and $\zeta_0=2.1$ in steps of $0.3$, whereas in panel~(b), we fix $\zeta_0=1.5$ but vary the initial charge $Z_0$. In both panels, the red, blue and green lines are assigned to the right ordinate and show the individual factors in Eq.~(\ref{eq:gammapropto}). Furthermore, the black curve marks the charge transfer amplitude as defined in Eq.~(\ref{eq:gammafinal}), with $d_\textup{w}=0.6a_0$ and $z_\textup{res}=-\sqrt{3}\,a_0$ as distance of resonance, where the amplitude reaches its maximum.}
\label{fig:gamma}
\end{figure}

%%\textcolor{red}{bevor die Abb diskutiert ist, muessten alle input groessen bekannt sein, auch $z_{res}$. Die Begründungen muessten also vorher erfolgen.}
In Fig.~\ref{fig:gamma}\!\!\!, we  show the charge transfer amplitude $\gamma_{k^*}(z)$ [orange curves] as defined in Eq.~(\ref{eq:gammapropto}) with the input of Eqs.~(\ref{eq:z}) [red curves] and (\ref{eq:omega}) [green curves] under variation of the parameters $\zeta_0$ [panel (a)] and $Z_0$ [panel (b)] and compare the obtained form as function of the ion-surface distance $z$ with the ansatz of Eq.~(\ref{eq:gammafinal}).
From both panels, we observe that
%% infer that
the charge transfer amplitude is indeed sharply peaked at some characteristic ion-surface distance and only for large values of $\zeta_0$ extends to considerably larger ion-surface distances. 
%%
%% \textcolor{red}{das müssten wir anpassen, da wir den peak ja selbst reingesteckt haben.}
%%
%% The existence of such a pronounced %%
Furthermore, we find that the position of the peak in $\gamma_{k^*}$ depends only very weakly on the ion's initial charge state.
%%, allows us to define a distance of resonance  (resonance point), $z_\textup{res}$, for which the probability for a GNF electron to perform a transition to the energy level $\epsilon_{k^*}$ of the ion has a clearly pronounced maximum. 
%%
The resonance point $z_\textup{res}=-\sqrt{3}a_0$ thereby describes the peak position reasonably well, cf.~the black vertical lines in Fig.~\ref{fig:gamma}\!\!\!, and thus determines the first open parameter in Eq.~(\ref{eq:gammafinal}). To also 
fix the parameter $d_\textup{w}$, we 
reproduce the width of the main peak in the orange curves for $\gamma_{k^*}(z)$ and set $d_\textup{w}=0.6a_0$. The resulting form of Eq.~(\ref{eq:gammafinal}) is hence the black curve in Fig.~\ref{fig:gamma}\!\!\!a and \ref{fig:gamma}\!\!\!b, which shows Eq.~(\ref{eq:gammafinal}) for $\gamma_0=1$ (scaled by a factor of $\times$ $0.65$).

%%and represents a reasonable form of the charge transfer amplitude,
%% particularly capturing the pronounced maximum predicted by Eq.~(\ref{eq:gammapropto}).

Finally, as we focus on resonant charge transfer processes, the point $z_\textup{res}=v_z t_\textup{res}$ also marks a specific GNF on-site energy, $E_{k^*}[\mathbf{S}(t_\textup{res})]=w[Z_0;\mathbf{R}_{k^*}-\mathbf{S}(t_\textup{res})]-W_\textup{f}$, which, in turn, provides a reasonable choice for the energy $\epsilon_{k^*}$. The resonance condition is given by
\begin{align}
\label{eq:resonancecondition}
    \epsilon_{k^*}\equiv E_{k^*}[\mathbf{S}(t_\textup{res})]=-\frac{Z_0W_0}{ \left(1+z_\textup{res}^2/a_0^2\right)^{1/2}}-W_\textup{f}\,,
\end{align}
and, for the resonance point $z_\textup{res}=-\sqrt{3}a_0$, the resonance condition~(\ref{eq:resonancecondition}) yields half the maximum induced field strength, i.e., $\epsilon_{k}=\epsilon=-Z_0W_0/2-W_\textup{f}$.
%% Moreover,
%and we use a Gaussian envelope for the charge transfer amplitude as a function of distance and time, respectively:
%\begin{align}
%\label{eq:gammafinal}
%    \gamma_k[{\mathbf S}(t)]=\gamma_0\exp\!\left(-\frac{(z(t)+z_\textup{res})^2}{2d_\textup{w}^2}\right)J_0\,.
%\end{align}
With the above choice of parameters, Eq.~(\ref{eq:gammafinal}) leaves open only one parameter, the prefactor $\gamma_0$, which will be used in Sec.~\ref{sec:4} to adapt the obtained results to xenon ion transmission experiments~\cite{Gruber2016}.

%In Fig.~\ref{fig:gamma}\!\!\!, we evaluate Eq.~(\ref{eq:gammapropto}) for reasonable assumptions on the functions $Z[\mathbf{S}(t)]$ and $\Omega_k[\mathbf{R}_k-\mathbf{S}(t)]$ and show the charge transfer amplitude as a function of the ion-surface distance $z(t)$. As reference site $k=k^*$, we choose one of the GNF's innermost lattice sites, which, with equal lateral distance $a_0$, are closest to the impact point of the ion, compare with Fig.~\ref{fig:gnf}\!\!\!a. 

%\textcolor{blue}{In the remainder of the article, we consider charge transfer only from lattice sites that belong to the GNF's innermost honeycomb (cf.~Fig.~\ref{fig:gnf}\!\!\!b) and, without loss of generality, use a more straightforward temporal (respectively spatial) shape for $\gamma_k[\mathbf{S}(t_\textup{res})]$, which is determined by two key parameters, the temporal (spacial) location of the amplitude's maximum and a characteristic temporal (spacial) width on which the charge transfer probability is nonzero. Compared to other rudimentary assumptions of the model, such as the usage of localized orbitals of the applied Hubbard model itself, we believe that this is no oversimplification, and the comparison with experimental data presented in Sec.~\ref{sec:4.1} supports this.}
%%
%% but is in line with the other such as the . 
%%
%\textcolor{blue}{Precisely, we} 

\section{Embedding self-energy approach}\label{sec:3}

To compute the electron dynamics resulting from Hamiltonian~(\ref{eq:ham}), we use the method of nonequilibrium Green functions (NEGF), see, e.g., Refs.~\cite{stefanucci13_cambridge,kadanoff-baym-book}. The central quantity, from which all relevant observables can be obtained, is the  one-particle nonequilibrium Green function of the graphene nanoflake,
\begin{align}
\label{eq:negf}
    G^{c}_{ij\sigma}(t,t')=-\frac{\textup{i}}{\hbar}\langle T_{\cal C}\,\hat{c}_{i\sigma}(t)\,\hat{c}^\dagger_{j\sigma}(t')\rangle_\textup{s}\,,
\end{align}
where $T_{\cal C}$ denotes time ordering on the Keldysh contour ${\cal C}$, and $\langle T_{\cal C}\dots\rangle_\textup{s}=\textup{tr}[T_{\cal C}\exp(A_\textup{s})\dots]/\textup{tr}[T_{\cal C}\exp(A_\textup{s})]$ indicates the ensemble average with $A_\textup{s}=-\frac{\textup{i}}{\hbar}\int_{\cal C}\textup{d}\bar{t}\,\hat{H}_\textup{s}(\bar{t})$. The density matrix and the site occupations of the GNF are recovered from the NEGF~(\ref{eq:negf}) in the limit of equal times,
\begin{align}
\langle\hat{\rho}_{ij\sigma}\rangle_\textup{s}(t)=\langle\hat{c}^\dagger_{i\sigma}\hat{c}_{j\sigma}\rangle_\textup{s}(t)&=-\textup{i}\hbar\,G^{c}_{ji\sigma}(t,t^+)\,,&\\
\langle\hat{n}_{i\sigma}\rangle_\textup{s}(t)&=-\textup{i}\hbar\,G^{c}_{ii\sigma}(t,t^+)\,,\nonumber
\end{align}
where the notation $t^+$ means, that the time $t^+$ is infinitesimally larger along the contour ${\cal C}$ than $t$. Further, the overall particle number is given by
\begin{align}
    \langle \hat{N}\rangle_\textup{s}(t)=-\textup{i}\hbar\sum_{i\sigma}G^{c}_{ii\sigma}(t,t^+)\,,
\end{align}
which needs to be multiplied by a factor of $4$ to obtain the total number of electrons in all four Hubbard bands of the present GNF model, cf.~Sec.~\ref{sec:2}. Similar to the NEGF of the graphene nanoflake, we define the charge transfer Green functions,
\begin{align}
\label{eq:ctnegf}
    G^{ca}_{ik\sigma}(t,t')&=-\frac{\textup{i}}{\hbar}\langle T_{\cal C}\,\hat{c}_{i
    \sigma}(t)\,\hat{a}^\dagger_{k\sigma}(t')\rangle_\textup{s}\,,&
    G^{ac}_{kj\sigma}(t,t')&=-\frac{\textup{i}}{\hbar}\langle T_{\cal C}\,\hat{a}_{k
    \sigma}(t)\,\hat{c}^\dagger_{j\sigma}(t')\rangle_\textup{s}\,,
\end{align}
which couple the GNF sites to the energy levels $\epsilon_k$ of the external ion (and vice versa).
Finally, the NEGF of the ion is
\begin{align}
\label{eq:ionnegf}
    G^{a}_{kk'\sigma}(t,t')&=-\frac{\textup{i}}{\hbar}\langle T_{\cal C}\,\hat{a}_{k
    \sigma}(t)\,\hat{a}^\dagger_{k'\sigma}(t')\rangle_\textup{s}\,,
\end{align}
and the superscripts indicate the type and arrangement of the creation and annihilation operators.

The equations of motion for the one-particle nonequilibrium Green functions~(\ref{eq:negf}), (\ref{eq:ctnegf}) and (\ref{eq:ionnegf}) are the Keldysh-Kadanoff-Baym equations~(KBE)~\cite{keldysh64,kadanoff-baym-book}:
\begin{align}
% cc
\label{eq:kbe1}
\sum_l\left[\textup{i}\hbar\,\partial_t \delta_{il}-H^{c}_{il\sigma}(t)\right]G^{cc}_{lj\sigma}(t,t')-\sum_{k\in M}\delta_{ik}\gamma_{k}[{\mathbf S}(t)]\,G^{ac}_{kj\sigma}(t,t')&=\delta_{\cal C}(t,t')\,\delta_{ij}+\sum_l\int_{\cal C}\textup{d}\bar{t}\,\Sigma^{c}_{il\sigma}(t,\bar{t})\, G^{c}_{lj\sigma}(\bar{t},t')\,,&\\
% ca
\label{eq:kbe2}
\sum_l\left[\textup{i}\hbar\,\partial_t \delta_{il}-H^{c}_{il\sigma}(t)\right]G^{ca}_{lk'\sigma}(t,t')-\sum_{k\in M}\delta_{ik}\gamma_{k}[{\mathbf S}(t)]\,G^{a}_{kk'\sigma}(t,t')&=0\,,&\\
% ac
\label{eq:kbe3}
\left[\textup{i}\hbar\,\partial_t -\epsilon_{k}(t)\right]G^{ac}_{kj\sigma}(t,t')-\sum_{k'\in M}\delta_{kk'}\gamma_{k'}[{\mathbf S}(t)]\,G^{c}_{k'j\sigma}(t,t')&=0\,,\\
% aa
\label{eq:kbe4}
\left[\textup{i}\hbar\,\partial_t -\epsilon_{k}(t)\right]G^{a}_{kk'\sigma}(t,t')-\sum_{k''\in M}\delta_{kk''}\gamma_{k''}[{\mathbf S}(t)]\,G^{ca}_{k''k'\sigma}(t,t')&=\delta_{\cal C}(t,t')\,\delta_{kk'}\,,
\end{align}
where $H^\textup{c}_{ij\sigma}(t)$ is the one-particle Hamiltonian of the graphene nanoflake,
\begin{align}
\label{eq:1pham}
    H^\textup{c}_{ij\sigma}(t)=\left\{
    \begin{array}{cc}
    W_i[Z_0;{\mathbf S}(t)]+E+U\left(\langle\hat{n}_{i\bar{\sigma}}\rangle-\frac{1}{2}\right)\,, & i=j\\[0.5pc]
    W_{\langle ij\rangle}[Z_0;{\mathbf S}(t)]-J\,,& ij=\langle ij\rangle\\[0.5pc]
    0\,,& \textup{otherwise}
    \end{array}
    \right.\,,
\end{align}
$\delta_{\cal C}(t,t')$ is the delta function on the Keldysh contour, and $\Sigma^{c}_{ij\sigma}(t,t')$ in Eq.~(\ref{eq:kbe1}) denotes the many-body self-energy~\cite{schluenzen_CondMat_2019} of the GNF. As mean-field (Hartree) contributions of the system (\ref{eq:ham}) are already contained in the one-particle Hamiltonian (\ref{eq:1pham}), the self-energy $\Sigma^{c}_{ij\sigma}(t,t')$ accounts for electron-electron correlations only. Moreover, we note that Eqs.~(\ref{eq:kbe1})-(\ref{eq:kbe4}) need to be accompanied by their adjoint equations, containing the time derivative with respect to $t'$.

In the following, we perform a formal decoupling of the GNF part of the Keldysh-Kadanoff-Baym equations~(\ref{eq:kbe1}--\ref{eq:kbe4}) from the ion part~\cite{stefanucci13_cambridge,bonitz-pssb2019}. To this end, we assume that the NEGF of the ion can be written as $G^a_{kk'\sigma}(t,t')=\delta_{kk'}g_{k\sigma}(t,t')$, where the nonequilibrium Green function $g_{k\sigma}(t,t')$ obeys the ideal (noninteracting) KBE
\begin{align}
\label{eq:kbeg}
    [\textup{i}\hbar\,\partial_t-\epsilon_k]g_{k\sigma}(t,t')=\delta_{\cal C}(t,t')\,,
\end{align}
which is solved by
\begin{align}
\label{eq:gepsilon}
    g_{k\sigma}(t,t')&=\frac{\textup{i}}{\hbar}[n_{k\sigma}-\theta_{\cal C}(t,t')]\exp\left(-\frac{\textup{i}}{\hbar}\epsilon_k(t-t')\right)\,,
\end{align}
where $\theta_{\cal C}(t,t')$ is the step function on the contour, and $n_{k\sigma}=\langle\hat{a}_{k\sigma}^\dagger\hat{a}_{k\sigma}\rangle_\textup{s}$ denotes the occupation of the energy level $\epsilon_k$. Using this ansatz, we find that the charge transfer Green function  can be expressed as
\begin{align}
\label{eq:ctnegfsolution}
G^{ac}_{kj\sigma}(t,t')=\int_{\cal C}\textup{d}\bar{t}\,g_{k\sigma}
(t,\bar{t})\,\gamma_k[{\mathbf S}(\bar{t})]\,G_{kj}^c(\bar{t},t')\,,
\end{align}
and insertion of (\ref{eq:ctnegfsolution}) into Eq.~(\ref{eq:kbe1}) leads to the closed equation of motion 
\begin{align}
\label{eq:kbefinal}
    \sum_l&\left[\textup{i}\hbar\,\partial_t \delta_{il}-H^{c}_{il\sigma}(t)\right]G^{c}_{lj\sigma}(t,t')
    =\delta_{\cal C}(t,t')\delta_{ij}+
    \sum_{l}\int_{\cal C}\textup{d}\bar{t}\,[\Sigma^{c}_{il\sigma}(t,\bar{t})+\Sigma^\textup{emb}_{il\sigma}(t,\bar{t})]\,G^{c}_{lj\sigma}(\bar{t},t')\,,
\end{align} 
where we have defined the \textit{embedding} self-energy
\begin{align}
\label{eq:sigmaemb}
    \Sigma_{ij\sigma}^\textup{emb}(t,t')=\left\{
    \begin{array}{cc}
    \gamma_i[{\mathbf S}(t)]
    \,g_{i\sigma}(t,t')\,\gamma_i[{\mathbf S}(t')]
    \,,& i=j\in M\\[0.5pc]
    0\,,&\textup{otherwise}
    \end{array}
    \right.\,.
\end{align}
Methodically, the additional embedding self-energy in the integration kernel on the r.h.s.~of Eq.~(\ref{eq:kbefinal}) incorporates the effects of resonant charge transfer into the GNF part~(\ref{eq:kbe1}) of the overall KBE and explicitly allows for temporal changes of the electron number $\langle\hat{N}\rangle_\textup{s}(t)$ in the graphene nanoflake. This enables us to characterize the loss of electrons from the GNF and, hence, the degree of neutralization of the external ion. Generally, the energy levels $\epsilon_k$ are  unoccupied before the ion interacts with the GNF, thus we set $n_{k\sigma}\equiv0$ in Eq.~(\ref{eq:gepsilon}). Moreover, as each energy level $k$ can capture one electron per spin $\sigma$, we introduce a number of ${\cal M}_\sigma=Z_0/2$ charge transfer channels $\hat{c}_{k\sigma}\rightarrow a^\dagger_{k\sigma}$ in the Hamiltonian, i.e., ${\cal M}_\sigma^\textup{s}={\cal M}_\sigma/4=Z_0/8$ transfer channels in each of the four Hubbard bands of Eq.~(\ref{eq:ham}). For the case of $Z_0$ being multiples of $8$, this implies the possibility of complete neutralization of the ion during its traversal of the nanoflake. Furthermore, in the considered scenario of normal incidence, charge transfer will occur predominantly from the GNF center, i.e., from lattice sites that are closest to the ion and the impact point. For this reason, we couple only sites of the nanoflake's central honeycomb. For the precise setup, which defines the set of indices $M$ in Eq.~(\ref{eq:sigmaemb}), see Fig.~\ref{fig:gnf}\!\!\!b.

The Keldysh-Kadanoff-Baym equation~(\ref{eq:kbefinal}) is the central equation which will be solved in the next section for various parameter sets under the neglect electron-electron correlations, i.e., for $\Sigma^{c}_{ij\sigma}\equiv0$ (mean field approximation). However, we note that, even for $\Sigma^{c}_{ij\sigma}\equiv0$, the remaining embedding part of the self-energy renders the equation of motion for $G^{c}_{ij\sigma}(t,t')$ still an integro-differential equation on the full Keldysh contour ${\cal C}$. Thus, the numerical effort is similar to solving a correlated problem, and the inclusion of electron correlations, e.g., in a second-order Born or T-matrix approximation, would be straightforward. To our advantage, the embedding self-energy~(\ref{eq:sigmaemb}) is local in space ($\Sigma_{ij\sigma}^\textup{emb}\propto\delta_{ij}\Sigma_{i\sigma}^\textup{emb}$), and, therefore, we use the auxiliary Hamiltonian approach developed in Ref.~\cite{balzer14_auxham} to efficiently solve the KBE.

\begin{figure}[b]
\begin{center}
\includegraphics[width=0.485\textwidth]{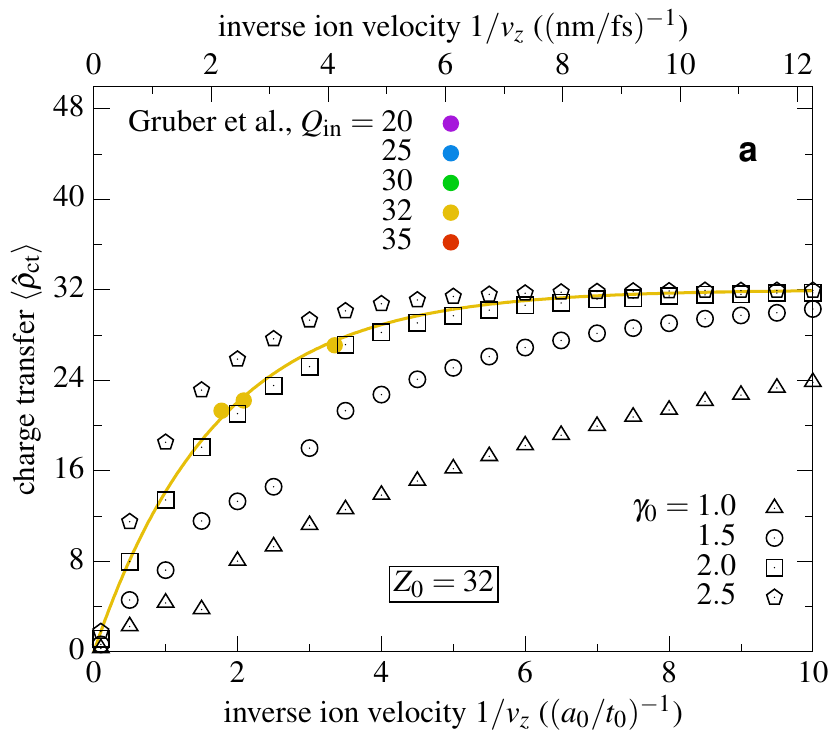}
\hspace{0.015\textwidth}
\includegraphics[width=0.485\textwidth]{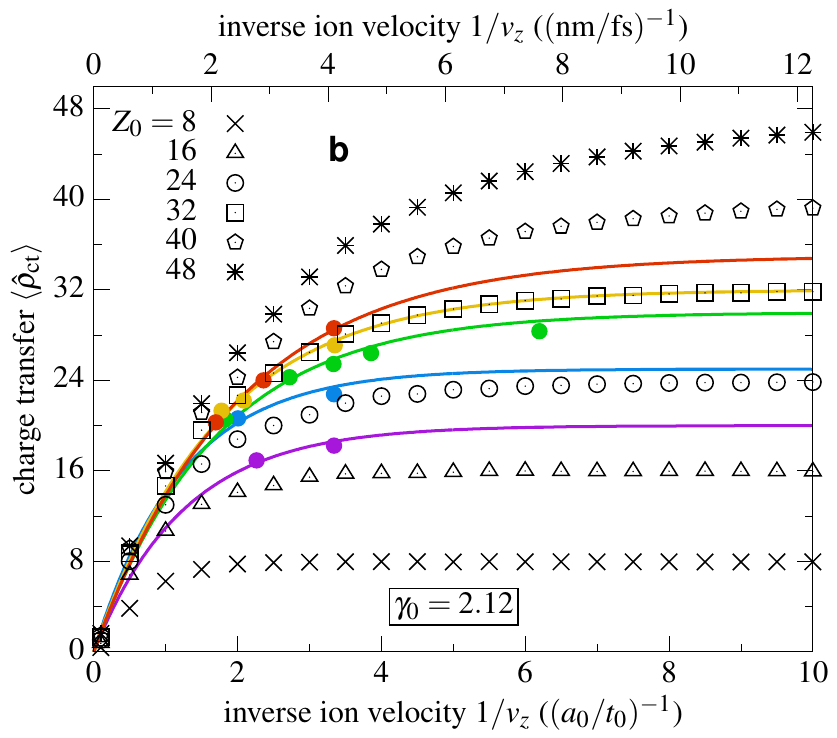}
\end{center}
\caption{Charge transfer $\langle \hat{\rho}_\textup{ct}\rangle$, black symbols, Eq.~(\ref{eq:rhoct}), as a function of the inverse ion velocity $v_z^{-1}$ obtained from the KBE~(\ref{eq:kbefinal}) for a graphene nanoflake with $L=24$ sites. Panel~(a): NEGF results for an ion with initial charge state $Z_0=32$ and different parameters $\gamma_0=1.0$, $1.5$, $2.0$ and $2.5$ in Eq.~(\ref{eq:gammafinal}). Panel~(b): NEGF results for different charge states $Z_0=8$, $16$, $24$, $32$, $40$, $48$ and $\gamma_0=2.12$. In both panels, the colored dots indicate experimental data of Gruber \textit{et al}., Ref.~\cite{Gruber2016}, for the transmission of  Xe$^{Q_\textup{in}+}$ ions through single-layer graphene, where $Q_\textup{in}=20$, $25$, $30$, $32$ and $35$, see legend in panel~(a). The colored solid lines are fits of the form (\ref{eq:qct-fit}) 
%$\langle\hat{\rho}_\textup{ct}\rangle=Q_\textup{in}(1-\textup{e}^{-a_0/(v_z\tau)})$ 
with the neutralization time constant $\tau$ fitted to the experimental data, assuming a continuous neutralization following an exponential function, compare with Ref.~\cite{Gruber2016}. %\textcolor{red}{gibt es eine erklärung für die Oszillationen der unteren 2 Kurven links? -> KB: Ich denke dies sind Resonanzeffekte, die sich aus der Tatsache ergeben, dass wir in der Tat nur ein effektives Energieniveau ankoppeln. Bei anderen Resonanzpunkten bzw. anderen Parameters in $\gamma(t)$ können diese Oszillationen auch durchaus noch stärker ausgeprägt sein (soweit ich die Parameter mal variiert habe).}
}
\label{fig:rhoct}
\end{figure}

\section{Results}\label{sec:4}

In this section, we solve the KBE~(\ref{eq:kbefinal}) with the charge transfer amplitude of Eq.~(\ref{eq:gammafinal}) and parameters $z_\textup{res}=-\sqrt{3}a_0$, $d_\textup{w}=0.6a_0$ and energies $\epsilon_{k}=\epsilon=-Z_0W_0/2-W_\textup{f}$ (setup of Fig.~\ref{fig:gnf}\!\!\!b as described in Sec.~\ref{sec:3}) for ions with initial charge states $Z_0=8$, $16$, $24$, $32$, $40$, $48$ and velocities $v_z$ in the range of 
$0.1$ and $100v_0$, i.e., with velocities between
%%
%% $0.1$ and $100\times0.815265157$\,nm/fs
%%
$8.2\times 10^4$ and $8.2\times10^7$\,m/s. Due to the extremely large and  time-dependent potential energy, which is induced inside the graphene nanoflake by the presence of the highly charged ion and which easily exceeds several hundred electron volts~\cite{Aumayr2003} (e.g., for $Z_0=32$, the maximum field strength on the central GNF lattice sites is given by 
%%
%% $-W_0Z_0=-324.4991939$\,eV
%%
$-W_0Z_0\approx-325$\,eV, all NEGF simulations are carried out on a fine ($t,t'$) time mesh. To resolve an extended time window of $[-75 t_0,75t_0]$, where the ion passes through the GNF layer at time $t=0$, we use $n_\textup{t}=12,000$ time steps, which corresponds to an integration step size and temporal resolution of $\Delta t/t_0=0.0125$. Initially, at time $t=-75t_0$, the graphene nanoflake is prepared in the half-filled Hartree ground state of the paramagnetic phase (the filling is $f=\frac{1}{L}\sum_{i\sigma}\langle\hat{n}_{i\sigma}\rangle_\textup{s}=0.5$ with $\langle\hat{n}_{i\uparrow}\rangle_\textup{s}=\langle\hat{n}_{i\downarrow}\rangle_\textup{s}\,\forall\,i$).

We emphasize again, that, for the described model setup and, following the self-energy approach of Sec.~\ref{sec:3}, the only remaining open parameter is the amplitude $\gamma_0$ of the charge transfer function (\ref{eq:gammafinal}). Moreover, the constraint of $Z_0$ being a multiple of $8$ is due to the restriction of the model Hamiltonian to four equivalent Hubbard bands ($4$ bands $\times$ $2$ spin-projections) in combination with the choice of considered charge transfer channels (cf.~Sec.~\ref{sec:3} and Fig.~\ref{fig:gnf}\!\!\!b).

\subsection{Charge transfer for different ion  velocities and charge states. Test of the model\label{sec:4.1}}

Our solutions of the KBE~(\ref{eq:kbefinal}) reveal that the number of electrons that are emitted from the graphene nanoflake sensitively depends on the interaction time between the HCI and the GNF [which is set by the ion velocity $v_z$] and on the ion's charge state $Z_0$ [which determines the strength of the local field inside the GNF lattice]. Figure~\ref{fig:rhoct}\!\!\! investigates, for a broad range of parameters, the overall transfer of charge onto the approaching ion, which is defined as 
\begin{align}
\label{eq:rhoct}
\langle\hat{\rho}_\textup{ct}\rangle=4\left(L-\langle \hat{N}\rangle_\textup{s} (t\rightarrow+\infty)\right)\,,
\end{align}
where $L$ denotes the number of lattice sites in the graphene nanoflake.

In Fig.~\ref{fig:rhoct}\!\!\!a, we consider a GNF with $L=24$ sites, an ion with $Z_0=32$ and vary the ion velocity $v_z$ as well as the amplitude $\gamma_0$. For small ion velocities and a sufficiently large value of $\gamma_0$, we observe that
the charge transfer (black symbols) allows  for a complete neutralization of the ion, i.e., $\langle\rho_\textup{ct}\rangle\sim Z_0$. On the other hand, for high ion velocities, the charge transfer rapidly decreases and tends towards zero, which is a direct consequence of the reduced  interaction time (at constant electron mobility) and ultimately leaves the electronic system undisturbed for $v_z^{-1}\rightarrow 0$. These two limiting cases are recovered independently of the choice of $\gamma_0$. 
On the other hand, at intermediate ion velocities,  the amount of electrons transferred to the ion and the dependence on $v_z$ sensitively depend on the choice of the charge transfer amplitude $\gamma_0$. A key question of the present work is, how accurately the simple model (\ref{eq:gammafinal}) describes the physical processes of the interaction of the charged projectile and the graphene cluster. Fortunately, there exists an extensive set of experimental data for the neutralization of 
Xe$^{Q_\textup{in}+}$ ions in single-layer graphene at similar impact parameters~\cite{Gruber2016}. The measured results for ions with an initial charge $Q_\textup{in}=32$ are shown by the orange dots in Fig.~\ref{fig:rhoct}\!\!\!a and show that the overall $v_z$-dependence is well captured by our model. Moreover, adjusting a single open parameter, $\gamma_0$, allows us to bring the two curves into complete agreement, for $\gamma\approx 2$.
The best fit is achieved for $\gamma_0=2.12$ which will be used in the following. 

%around $2.0$, we can map the NEGF results onto data of transmission experiments, which have been performed on single-layer graphene, i.e., basically on the same material, with Xe$^{Q_\textup{in}+}$ ions at similar impact parameters~\cite{Gruber2016}, cf.~the orange dots in Fig.~\ref{fig:rhoct}\!\!\!a being associated with the initial charge state $Q_\textup{in}=32$.

\begin{figure}[b]
\begin{center}
\includegraphics[width=0.33\textwidth]{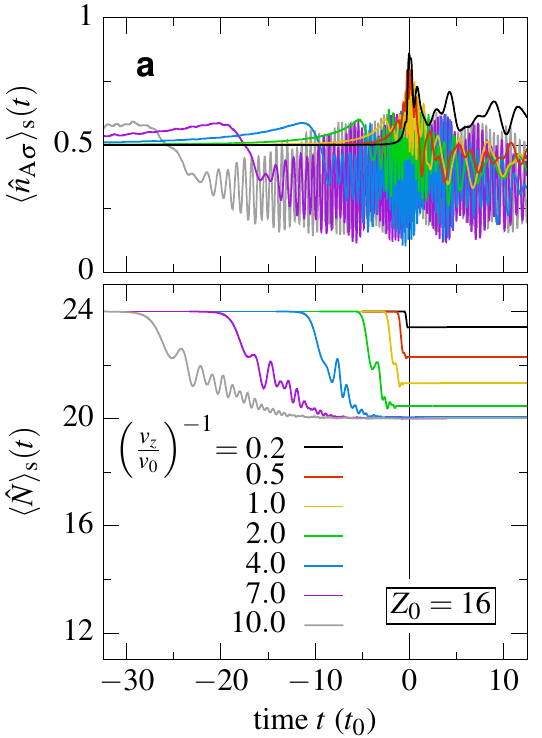}
\includegraphics[width=0.33\textwidth]{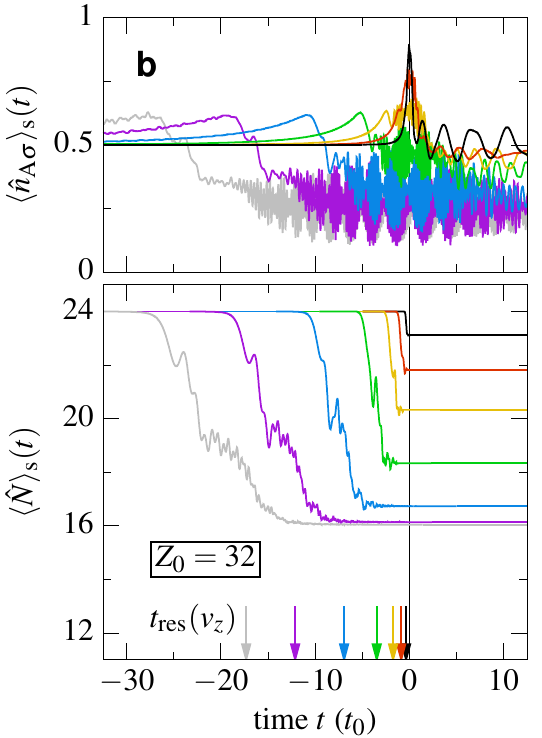}
\includegraphics[width=0.33\textwidth]{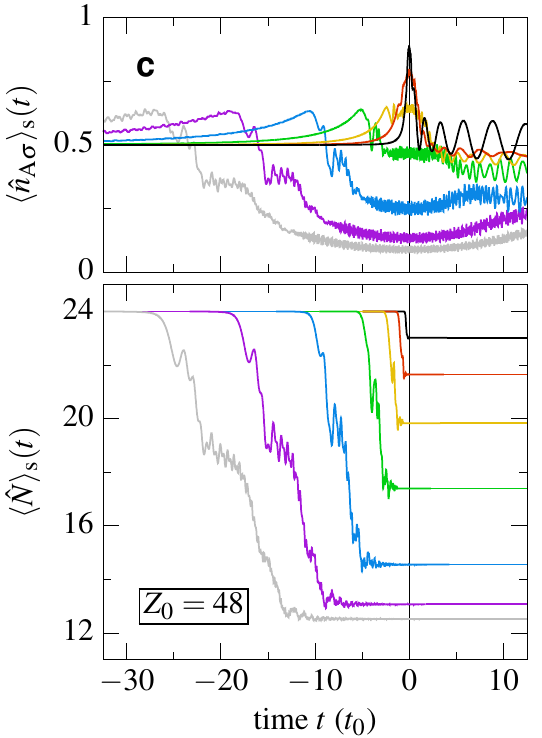}
\end{center}
\caption{Time evolution of the overall particle number $\langle\hat{N}\rangle_\textup{s}(t)$ and the site occupation $\langle\hat{n}_{0\sigma}\rangle_\textup{s}(t)$ [central site of the GNF, cf.~site index $\textup{A}$ in Fig.~\ref{fig:gnf}\!\!\!a] during ion impact for a graphene nanoflake with $L=24$ sites and (a)~an initial ion charge state $Z_0=16$, (b)~$Z_0=32$ and (c)~$Z_0=48$. The different colors refer to NEGF results for different ion velocities that are given in the legend of the left figure.
%$(v_z/v_0)^{-1}=0.2$, $0.5$, $1.0$, $2.0$, $4.0$, %$7.0$ and $10.0$. 
All other parameters are as in Fig.~\ref{fig:rhoct}\!\!\!b.  %\textcolor{blue}{
In panel (b), the arrows indicate the time $t_\textup{res}=z_\textup{res}/v_z$, when the ion passes through the resonance point $z_\textup{res}=-\sqrt{3}a_0$, in front of the plane of the graphene nanoflake.
%} 
%\textcolor{red}{waere interessant zu zeigen, wie weit das Ion weg ist, wenn oben der erste peak auftritt, bzw. wo der Resonanzpunkt ist (zu welcher Zeit erreicht)}
}
\label{fig:density}
\end{figure}

The next question is whether this good agreement is just a coincidence for a single charge state. To answer this, we display in Fig.~\ref{fig:rhoct}\!\!\!b,
all experimental data available from Ref.~\cite{Gruber2016}, i.e., in addition to $Q_\textup{in}=32$, also the charge transfer for $Q_\textup{in}=20$, $25$, $30$, and $35$, cf. the colored dots and the fits (lines of the same color). These data are compared to our simulations for different charge states that are multiples of eight (see legend in the figure) where we now fix $\gamma_0=2.12$ that was obtained from the case $Q_\textup{in}=32$ (left figure). The striking result is that this single parameter value allows us to reproduce the experimental curves for all available charge states, compare, e.g., the blue line for $Q_\textup{in}=25$ with the simulations for $Z_0=24$ (black circles). 
For other values agreement can be established by interpolating between the available charge values. This gives strong support for the present simple model (\ref{eq:gammafinal}). Moreover, this allows us also to 
predict the charge transfer behavior for small impact velocities and additional charge states for which no experimental data are available.
%
%to obtain a good agreement with the experimental data for Xe$^{32+}$ and the corresponding fit (cf.~the orange line). Moreover, we show NEGF results for the remaining ion charge states considered in the present model and  
%

This good agreement with the experiments gives us confidence that our model, combined with our KBE dynamics, captures the correct physics. Generally, we find that ions with a lower initial charge state get fully neutralized already at smaller inverse ion velocities and thus have smaller neutralization time constants $\tau$, cf.~also Ref.~\cite{Gruber2016}. 
From the figure and, using the fit formula 
\begin{align}
\langle\hat{\rho}_\textup{ct}\rangle=Q_\textup{in}\left(1-\textup{e}^{-a_0/(v_z\tau)}\right),    
\label{eq:qct-fit}
\end{align}
we furthermore extract neutralization time constants in the range of $\sim\!0.7\,t_0$ (for $Z_0=8$) to $\sim\!2.6\,t_0$ (for $Z_0=48$), with respect to the ion travel distance of one lattice spacing (carbon-carbon bond length $a_0$). %\textcolor{red}{wie wurde $\tau$ bestimmt?}

\subsection{Density response and  finite size effects\label{sec:4.2}}
After having verified the good quality of our model, in this section, we analyze more details of the charge transfer. In particular, we investigate the time-resolved electronic response of the graphene nanoflake during impact of the HCI and perform simulations for GNFs of different size.

\begin{figure}[b]
\begin{center}
\includegraphics[width=0.485\textwidth]{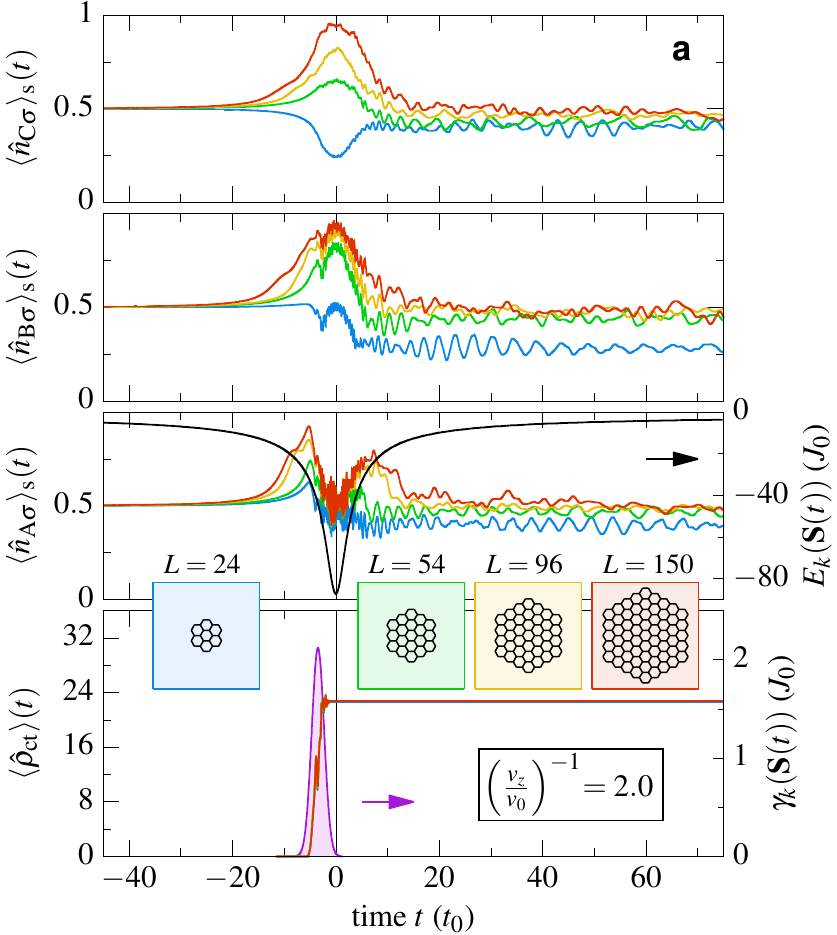}
\hspace{0.015\textwidth}
\includegraphics[width=0.485\textwidth]{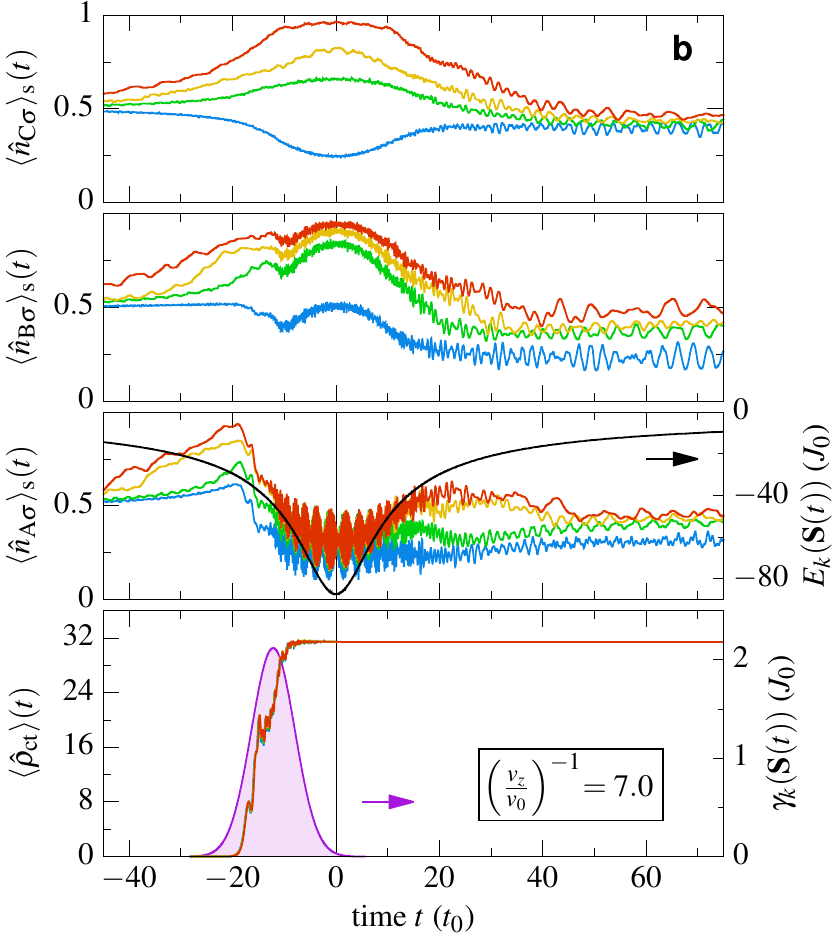}
\end{center}
\caption{Electronic response to ion impact of different hexagonal hollow-centred GNFs with $L=24$ (blue curves), $54$ (green), $96$ (orange) and $150$ (red) lattice sites. Ion parameters:~Initial charge state $Z_0=32$ and velocity $v_z$ as indicated. \textbf{Upper three panels}: time-dependent occupations $\langle\hat{n}_{i\sigma}\rangle_\textup{s}(t)$ for sites $i=\textup{A},\textup{B},\textup{C}$ as defined in Fig.~\ref{fig:gnf}\!\!\!a. \textbf{Bottom panel}: time-resolved charge transfer $\langle\hat{\rho}_\textup{ct}\rangle(t)$, Eq.~(\ref{eq:rhoct}), as well as the charge transfer function $\gamma_k[{\mathbf S}(t)]$ (violet peak). Furthermore, the black line in the third row indicates the time-dependent potential energy $E_k[{\mathbf S}(t)]$ on site $\textup{A}$ and all other sites of the central honeycomb (right ordinate).}
\label{fig:density2}
\end{figure}

In Fig.~\ref{fig:density}\!\!\!a-c, we monitor the total particle number in a single-band, $\langle\hat{N}\rangle_\textup{s}(t)$, (lower panels) and the mean occupation of the central site, $\langle\hat{n}_{\textup{A}\sigma}\rangle_\textup{s}(t)$, (upper panels) around the time of impact ($t=0$) for a GNF with $L=24$ sites and three different values of $Z_0$ at various ion velocities. Generally, the electrons are attracted by the approaching ion and thus move inside the graphene nanoflake, i.e., along the edges of the honeycomb lattice, towards the impact point and accumulate in the GNF center (cf.~also Ref.~\cite{Gruber2016}). This explains the increase of $\langle\hat{n}_{\textup{A}\sigma}\rangle_\textup{s}$, for $t<0$, before the onset of electron emission given by the decline of $\langle\hat{N}\rangle_\textup{s}$. The accumulation of electrons in the GNF center is thereby more pronounced for large ion velocities, as in this case less electrons are simultaneously emitted from the central honeycomb. Particularly for $v_z^{-1}<1.0$, this even leads to a  clearly enhanced double occupation $\langle\hat{n}_{\textup{A}\uparrow}\hat{n}_{\textup{A}\downarrow}\rangle\approx\langle\hat{n}_{\textup{A}\uparrow}\rangle\langle\hat{n}_{\textup{A}\downarrow}\rangle$ (Hartree approximation) in the graphene nanoflake and, thus, potentially to the formation of doublons~\cite{Balzer_PRL_2018}. Moreover, the emission of electrons leads to the emergence of fast Bloch oscillations of the central density, the amplitude (frequency) of which decreases (increases) with the field strength generated by the ion. This is seen in the upper panels of Fig.~\ref{fig:density}\!\!\! a to c where the ion charge is varied form 16 to 48. 

For $t>0$, the ion quickly departs from the nanoflake. Since in the present model no further significant charge transfer occurs, the total particle number remains constant. At the same time, the density on the central site finally starts to relax in an oscillatory manner where details depend on the remaining filling of the GNF. Also, the period of the oscillation increases with increasing distance of the ion due to the decay of its Coulomb field.
The equilibration process is considered in some more detail in Fig.~\ref{fig:density2}\!\!\!, which shows the time evolution of the
electron density along a path $\overline{\textup{A}\textup{B}\textup{C}}$ of adjacent sites (see Fig.~\ref{fig:gnf}\!\!\!a) away from the GNF center and discusses the influence of the cluster size $L$. On the innermost site $\textup{A}$, from which electron emission occurs, we observe that, independently of the ion velocity, the time-dependent occupation around the time of ion impact converges for $L\gtrsim96$, whereas for times essentially before and after the impact as well as on sites $\textup{B}$ and $\textup{C}$ the dynamics still depends on the cluster size. On sites $\textup{B}$ and $\textup{C}$, we clearly see the accumulation of electrons during the impact of the ion, particularly for cluster sizes $L\geq54$, where the maximum site occupation increases with the cluster size and approaches values $\langle\hat{n}_{\textup{C}\sigma}\rangle_\textup{s}\gtrsim0.9$, for the largest nanoflake, cf.~in particular the top panel in Fig.~\ref{fig:density2}\!\!\!b, where the site occupation remains large for several femtoseconds. For $L=24$, we note that site $\textup{C}$ is located at the edge of the nanoflake, and electron transfer towards the center is not compensated from more distant sites. Therefore, site C exhibits a density decrease, in striking contrast to the other sites. Moreover, due to the electron capture and the finite nature of the graphene nanoflakes, the ion always leaves behind nanoclusters with a filling $f<0.5$. This deviation from the half-filled case is of course most pronounced for the smallest GNF containing the smallest amount of electrons, see the blue curves which, for $t\rightarrow\infty$, indicate final occupations considerably less than $0.5$.

Finally, the bottom panels of Fig.~\ref{fig:density2}\!\!\! show that the overall charge transfer does in the present model not depend significantly on the size of the graphene nanocluster (all curves are virtually on top of each other). We note, that this is the case even though the occupations on site $\textup{A}$ clearly differ for the different nanoflakes directly before the onset of the charge transfer.
\\
%%\textcolor{red}{KB: Ja, die Information ist im Prinzip vorhanden. Hast Du die Nichtdiagonal-Elemente der DM? Die Elektrische Stromdichte bzw. Stromstärke zwischen A-B und B-C wäre noch interessanter als die Dichten auf den sites A, B, C, daraus könnte man auch eine Leitfähigkeit abschätzen und mit der von Graphen vergleichen, wenn man das Potential des Ions verwendet. Wäre evtl. etwas fuer ein follow up.}

\section{Discussion and conclusions}\label{sec:5}

In summary, we have studied the
time-dependent neutralization dynamics of highly charged ions which resonantly capture electrons during transmission through a two-dimensional atomic monolayer.
This was done by extending our recently developed
Nonequilibrium Green functions-Ehrenfest dynamics  approach \cite{balzer-schluenzen_PRB_2016,Balzer_PRL_2018} to include charge transfer processes. As test systems we have chosen finite, hexagonal graphene nanoflakes. Our primary focus has been on the electronic response of the GNF layer, while the treatment of the ion was confined to the occupation of excited states with an effective energy $\epsilon_k=\epsilon$ that correspond to a transient configuration of a ``hollow atom (ion)''. From the information about the occupied level $\epsilon_k$ we reconstructed the one-particle NEGF of the ion [Eq.~(\ref{eq:ionnegf})] which allowed us to compute the two-time embedding self-energy $\Sigma^\textup{emb}_{ij\sigma}(t,t')$ [Eq.~(\ref{eq:sigmaemb})]. 
Due to the appearance of the embedding self-energy in the integration kernel of the Keldysh-Kandaoff-Baym equation~(\ref{eq:kbefinal}), the number of electrons in the graphene nanoflake is no longer conserved, even for a conserving approximation of the regular many-body self-energy $\Sigma^c_{ij\sigma}(t,t')$, and thus, by construction, the change of particle number corresponds to the number of electrons emitted from the GNF. This procedure resembles the embedding schemes that were used before in NEGF calculations of transient photoabsorption~\cite{Perfetto_2015_PRA,Perfetto_2018_CHEERS} or of the transient dynamics in quantum transport setups~\cite{Myohanen_2009}. The NEGF simulations were combined with an analytical parametrization of the charge transfer amplitude. Comparison to available experimental data for highly charged ions revealed excellent agreement giving strong support for our approach.

Let us discuss limitations and possible future improvements.
The first limitation of our approach is that the initially formed hollow ion state does not undergo further de-excitation  as a function of time. These processes might influence the neutralization as well as the charge transfer. For fast impact velocities, and hence small GNF-ion interaction times, such de-excitation effects should, however, not essentially affect the emission of electrons from the graphene nanoflake. Secondly, we have assumed that the potential energy induced inside the GNF [Eqs.~(\ref{eq:wpotential}) and (\ref{eq:wi})] only depends on the HCI's initial charge $Z_0$ during the whole interaction process, although the capture of electrons modifies the charge state as function of time. This effect could be incorporated into the model via an effective, time-dependent charge state $Z(t)$, which is constructed from the number of electrons emitted from the nanoflake. For the equation of motion of the nanoflake's NEGF, such a feedback loop would, however, necessitate an iterative solution of the problem. It remains to be checked whether the KBE will converge in such a procedure. On the other hand, we are confident that the current ansatz is again justified for sufficiently fast ions, where electrons move into energetically high-lying and loosely bound hollow state wave packets, with which the definition of a clear ion charge state is delicate anyway. Furthermore, the excellent agreement of our NEGF results with the experimental data for a reasonably motivated choice of the charge transfer amplitude [Sec.~\ref{sec:2.2}] gives us confidence that the approach includes most of the relevant physics. A thorough comparison of the results to predictions obtained from a classical over-the barrier model~\cite{Burgdoerfer_1991} is left for future work. The same concerns the extension of the simulations to include charge transfer from  lattices sites outside the innermost honeycomb ring. Moreover, an interesting topic is to investigate the influence of the charge transfer on the stopping power and the modification of the charge transfer by the initial acceleration of the ion \cite{balzer-schluenzen_PRB_2016}.

The main advantage of the embedding self-energy approach is that the numerical effort is not substantially larger compared to a NEGF simulation for the closed, isolated graphene nanoflake system and that it requires only minor adjustments of existing NEGF codes. In addition, the calculations can straightforwardly be extended beyond the mean-field level to incorporate electron-electron correlations, e.g., using the (local) second-order Born approximation, as in Ref.~\cite{balzer-schluenzen_PRB_2016}. More advanced self-energies (including GW or T-matrix), larger systems and more complex and realistic Hamiltonians should become accessible in conjunction with the current development of more powerful and efficient KBE solvers~\cite{Schluenzen_PRL2020,joost_PRB_2020} based on the generalized Kadanoff-Baym ansatz. Finally, despite the above mentioned limitations, the approach offers further interesting applications such as the investigation of charge transfer for electron- or hole-doped graphene nanoflakes or the study of correlation effects and doublon formation for systems with substantially larger Coulomb interaction strengths and, hence, pronounced band gaps.

%% Potential extensions:
%% -- adaptation of the NEGF of the ion, multiple energy levels? 
%% -- increase of the interaction region (here limited to the central GNF honeycomb)

\section*{Acknowledgments}
We acknowledge Lasse Wulff's contributions to the code development at the initial stage. Further, we acknowledge computing time at the Computing Centre of Kiel University. 

%\bibliography{kb-ref}

%% \begin{thebibliography}
%% \bibitem[a] bc.
%% \end{thebibliography}

\providecommand{\url}[1]{\texttt{#1}}
\providecommand{\urlprefix}{}
\providecommand{\foreignlanguage}[2]{#2}
\providecommand{\Capitalize}[1]{\uppercase{#1}}
\providecommand{\capitalize}[1]{\expandafter\Capitalize#1}
\providecommand{\bibliographycite}[1]{\cite{#1}}
\providecommand{\bbland}{and}
\providecommand{\bblchap}{chap.}
\providecommand{\bblchapter}{chapter}
\providecommand{\bbletal}{et~al.}
\providecommand{\bbleditors}{editors}
\providecommand{\bbleds}{eds: }
\providecommand{\bbleditor}{editor}
\providecommand{\bbled}{ed.}
\providecommand{\bbledition}{edition}
\providecommand{\bbledn}{ed.}
\providecommand{\bbleidp}{page}
\providecommand{\bbleidpp}{pages}
\providecommand{\bblerratum}{erratum}
\providecommand{\bblin}{in}
\providecommand{\bblmthesis}{Master's thesis}
\providecommand{\bblno}{no.}
\providecommand{\bblnumber}{number}
\providecommand{\bblof}{of}
\providecommand{\bblpage}{page}
\providecommand{\bblpages}{pages}
\providecommand{\bblp}{p}
\providecommand{\bblphdthesis}{Ph.D. thesis}
\providecommand{\bblpp}{pp}
\providecommand{\bbltechrep}{}
\providecommand{\bbltechreport}{Technical Report}
\providecommand{\bblvolume}{volume}
\providecommand{\bblvol}{Vol.}
\providecommand{\bbljan}{January}
\providecommand{\bblfeb}{February}
\providecommand{\bblmar}{March}
\providecommand{\bblapr}{April}
\providecommand{\bblmay}{May}
\providecommand{\bbljun}{June}
\providecommand{\bbljul}{July}
\providecommand{\bblaug}{August}
\providecommand{\bblsep}{September}
\providecommand{\bbloct}{October}
\providecommand{\bblnov}{November}
\providecommand{\bbldec}{December}
\providecommand{\bblfirst}{First}
\providecommand{\bblfirsto}{1st}
\providecommand{\bblsecond}{Second}
\providecommand{\bblsecondo}{2nd}
\providecommand{\bblthird}{Third}
\providecommand{\bblthirdo}{3rd}
\providecommand{\bblfourth}{Fourth}
\providecommand{\bblfourtho}{4th}
\providecommand{\bblfifth}{Fifth}
\providecommand{\bblfiftho}{5th}
\providecommand{\bblst}{st}
\providecommand{\bblnd}{nd}
\providecommand{\bblrd}{rd}
\providecommand{\bblth}{th}

\end{document}